# Exciton Localization on Ru-based Photosensitizers Induced by Binding to Lipid Membranes

Pedro A. Sánchez-Murcia,* Juan José Nogueira and Leticia González*

*Institute of Theoretical Chemistry, Faculty of Chemistry, University of Vienna, Währinger Str. 17 A-1090 Vienna, Austria*

**Corresponding Author**

* pedro.murcia@univie.ac.at

* leticia.gonzalez@univie.ac.at

# ABSTRACT

The characterization of electronic properties of metal complexes embedded in membrane environments is of paramount importance to develop efficient photosensitizers in optogenetic applications. Molecular dynamics and QM/MM simulations together with quantitative wavefunction analysis reveal an unforeseen difference between the exciton formed upon excitation of [Ru(bpy)$_2$(bpy-C17)]$^{2+}$ embedded in a lipid bilayer and in water, despite the media does not influence the charge-transfer character and the excitation energy of the absorption spectra. In water the exciton is mainly delocalized over two ligands and the presence of non-polar substituents induces directionality of the electronic transition. Instead, when the photosensitizer is embedded into a lipid membrane, the exciton size diminishes and is more localized at one bypyridyl site due to electrostatic interactions with the positively charged surface of the bilayer. These differences show that the electronic structure of metal complexes can be controlled through the binding to external species, underscoring the crucial role of the environment in directing the electronic flow upon excitation and thus helping rational tuning of optogenetic agents.

**TOC GRAPHICS**

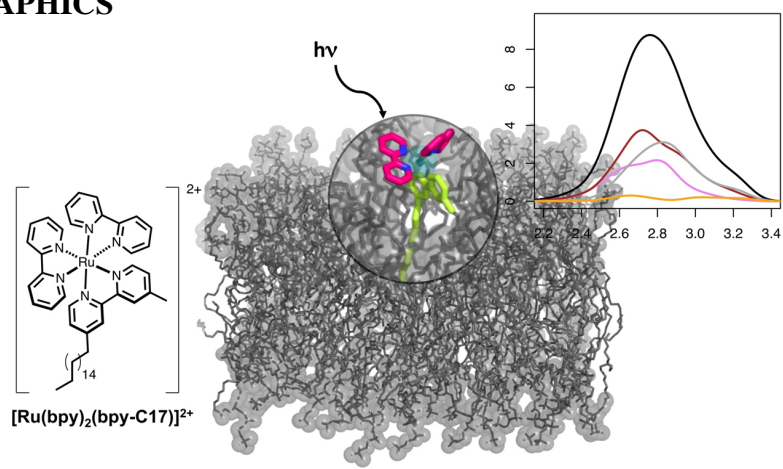

Ru(II)-based complexes are powerful photosensitizers with attractive biological and medical applications,[1-3] including optogenetic cellular control. Recently, it has been shown that the complex [Ru(bpy)$_2$(bpy-C17)]$^{2+}$ (Figure 1A) is able to insert into cell membranes and induce alterations in the cell electrical activity[4] upon illumination with light. Given the potential of optogenetic reagents, the importance of understanding the behavior of generated excited species on lipid membranes cannot be understated. However, an unequivocal electronic characterization of metal complexes excited states is a very challenging task, and even more in a biological environment. The excited electron can acquired different degrees of delocalization (also termed the exciton size) that not only depend on the chemical structure of the complex[5-6] but also on the time scale of the experimental measurement.[7-8] The need of considering vibrational motion and environmental effects in the characterization of the electronic structure of the chromophore poses a severe challenge to theory.[9-11] This complexity is very well represented by the archetype photosensitizer [Ru(bpy)$_3$]$^{2+}$.[12] After light absorption, the complex populates a singlet metal-to-ligand charge-transfer state ($^1$MLCT), which efficiently goes to a triplet $^3$MLCT state by intersystem crossing in a ultrafast manner (~25 fs).[8, 13-14] The long lived $^3$MLCT state (~600-800 ns) is well characterized as it has been found that the excited electron localizes on only one of the bipyridine units (Figure 1B).[7, 15] In contrast, different degrees of delocalization have been reported for the initially populated $^1$MLCT state,[7-8, 12, 14, 16-17] likely, due to its very short lifetime (<30 fs), which makes it hard to analyze.[18]

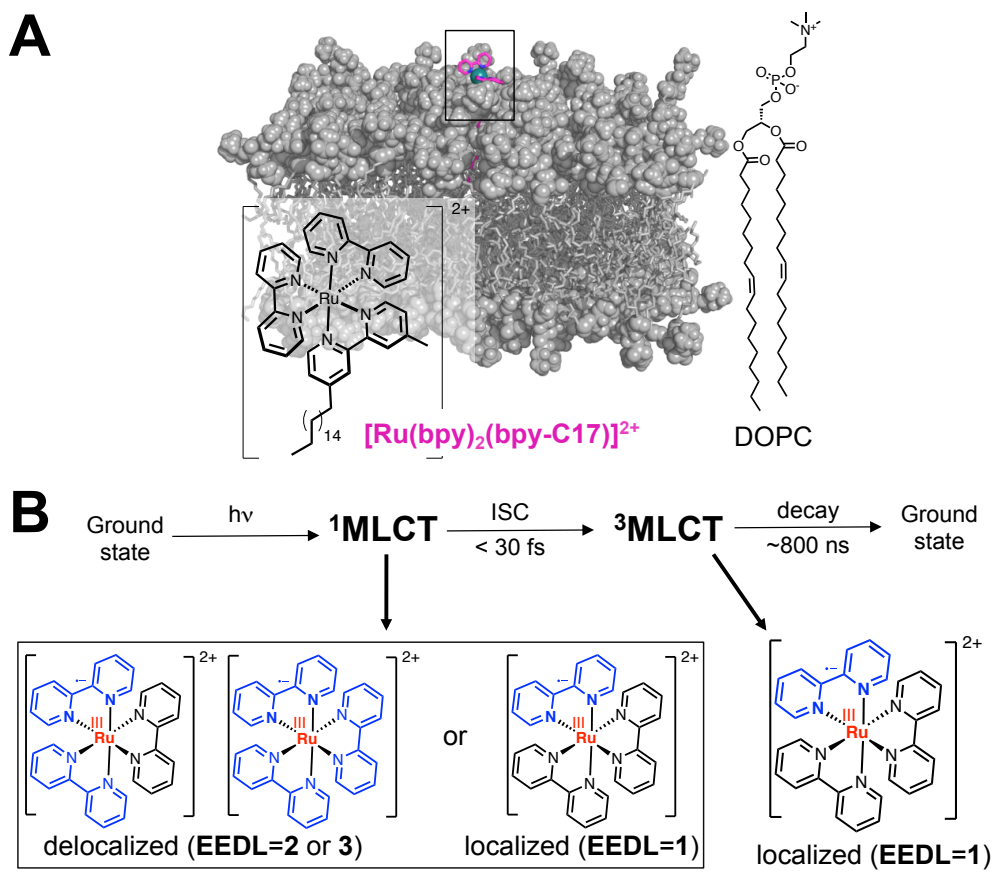

**Figure 1**. (A) Chemical structure of [Ru(bpy)$_2$(bpy-C17)]$^{2+}$ and dioleoyl phosphatidylcholine (DOPC) lipid chains that compose the membrane. (B) Schematic representation of the photophysics of [Ru(bpy)$_3$]$^{2+}$ indicating the possible excited electron delocalization lengths (EEDL) for the singlet and triplet electronic states.[12]

Here we report the first exciton quantification in a Ru-based photosensitizer inside a lipid membrane. As an example, we take [Ru(bpy)$_2$(bpy-C17)]$^{2+}$, as this Ru diimine complex has been the first to use in optogenetics.[4] We unambiguously characterize the exciton length of [Ru(bpy)$_2$(bpy-C17)]$^{2+}$ formed immediately after electronic excitation bound to a lipid bilayer and for comparison in aqueous solution. We show that this initial exciton distribution is different in both environments after light absorption. This finding is valuable for the rational design of novel Ru-based devices by chemical substitutions and by binding to external agents.

Our initial step is the incorporation of the Ru-based photoactivatable molecule into a lipid membrane (Figure 1A). The complex was inserted in the membrane by means of umbrella sampling classical molecular dynamics (US-MD), in which the whole system was described by a force field.[19-23] A dioleoyl phosphatidylcholine (DOPC) membrane was chosen as model because PC is the major component of biological membranes[24] and ubiquitously used in simulations.[25] Subsequent hybrid quantum mechanics/molecular mechanics (QM/MM) MD simulations, where the metal complex is described by density functional theory (DFT) with the B3LYP functional,[26] were carried out to refine the classical geometries. For comparison, the Ru complex was also simulated in aqueous solution with the same procedure, i.e. classical MD and subsequent QM/MM MD simulations (see Section S1 of SI for more details). Our simulations show that [Ru(bpy)$_2$(bpy-C17)]$^{2+}$ partially inserts the aliphatic chain into the lipid membrane, leaving the organometallic core exposed to the solvent and interacting with the surrounding PC moieties (see Figure 2B). The PC groups of the membrane are reorganized to favor electrostatic interactions with [Ru(bpy)$_2$(bpy-C17)]$^{2+}$. Whereas the negatively charged oxygen atoms of the phosphates and ester groups point towards the ruthenium coordination sphere, which is positively charged, the ammonium groups are projected outwards to avoid electrostatic repulsion. Thus, the cationic metal complex is embedded into a pocket that possesses a negative electrostatic potential, represented by the blue mesh in Figure 2A.

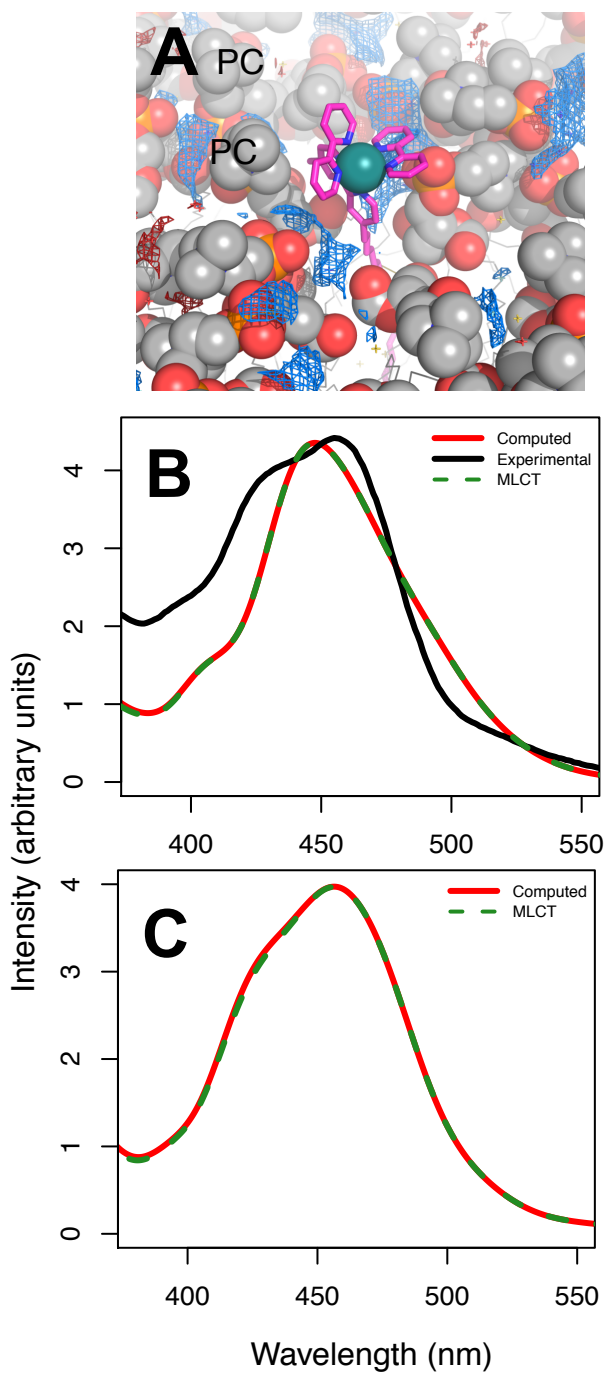

**Figure 2.** (A) Representative snapshot taken from the QM/MM MD trajectory showing the binding mode of [Ru(bpy)$_2$(bpy-C17)]$^{2+}$ into the membrane. Color code: the regions with negative electrostatic potential are represented by the blue mesh, oxygen atoms of the phosphate in red, the trimethyl groups of PC in grey. (B) Experimental[4] and computed absorption spectrum of [Ru(bpy)$_2$(bpy-C17)]$^{2+}$ in water and contribution of metal-to-ligand charge-transfer (MLCT) transitions. (C) Computed absorption spectrum of [Ru(bpy)$_2$(bpy-C17)]$^{2+}$ embedded into the membrane and contribution of metal-to-ligand charge-transfer (MLCT) transitions.

As a next step, we computed the UV-vis absorption spectra of [Ru(bpy)$_2$(bpy-C17)]$^{2+}$ in aqueous solution and into the lipid bilayer in order to evaluate the effect of the environment. To this aim, 50 snapshots from the QM/MM MD simulations for each environment were selected, for which the electronic excitation energies of the 20 lowest-lying singlet states were calculated using an electrostatic-embedding QM/MM scheme. As before, the metal complex is described by the B3LYP functional[26], in its time-dependent fashion (TD-DFT), and the environment is described by force fields.[22] Further computational details and a discussion about the choice of the electronic-structure method can be found in Section S2 of the SI. The obtained spectrum in water (Figure 2B) shows very good agreement with the experimental one,[4] validating the employed computational protocol. Noticeably, the shapes of the absorption bands of [Ru(bpy)$_2$(bpy-C17)]$^{2+}$ in water and inside the membrane are very similar, with a mere red-shift of 34 meV when going from water to the bilayer (Figure 2C). Such small shift agrees with the fact that experimentally this absorption band in [Ru(bpy)$_3$]$^{2+}$ is also vaguely sensitive to the solvent.[27] For comparison, the absorption spectrum of [Ru(bpy)$_2$(bpy-C17)]$^{2+}$ was also computed in a pure non-polar environment, by completely integrating the complex into the lipid membrane; as expected, no significant changes in the shape were observed (see Section S2 of the SI). Figures 2B,C also show the contribution of MLCT character to the UV/vis absorption spectrum in both media, calculated by an analysis of the transition density matrix[28-29] (section S3 of the SI for further details). As it can be seen, again, no significant differences were found between water and the lipid membrane, indicating that the media does not influence the character of the electronic absorption, which is completely dominated by MLCT transitions. One would be therefore left to conclude that a water environment is a reasonable description of the photosensitizer in the lipid membrane –justifying why many studies of compounds in biological environments begin with a characterization of the relevant species in water, as a first and reasonable approximation. This equivalency, as we shall demonstrate in the following, is in many cases an apparent fallacy.

Disgruntled by the similar effect that the lipid and the aqueous environments seems to have on the UV/Vis absorption spectrum of [Ru(bpy)$_2$(bpy-C17)]$^{2+}$, we analyzed the excited electron localization paying attention to the chemical structure of the complex. To do this, the absorption band was decomposed into different excited-electron delocalization length (EEDL) contributions. EEDL is defined here as the number of fragments over which an exciton is delocalized.[28-29] In particular, we divided [Ru(bpy)$_2$(bpy-C17)]$^{2+}$ into four different fragments, the three ligands and the core Ru atom.[30]

To our surprise, and contrary to the spectral shape and excited state character, the EEDL descriptor, i.e. the delocalization of the exciton, shows significantly changes depending on the environment (Figures 3A,B). In aqueous solution, most of electronic transitions are delocalized over two ligands (61%), with smaller contributions within one (11%) and three ligands (27%). These results are in line with previous simulations where the excited electron was found to be delocalized mainly on one or two bipyridine ligands due to the break of the symmetry because of the presence of solvent molecules.[31] In contrast, when the metal complex is embedded into the bilayer, the electronic excitation localizes: the contribution of excitations located on only one ligand increases from 11% to 25%, delocalization over two ligands decreases from 61% to 40%, and EEDLs over three ligands and four fragments barely change. Therefore, when going from aqueous solution to the lipid bilayer the chromophore undergoes a net localization of the excited electron. The electrostatic interactions between the chromophore and the lipid membrane are responsible for this net electronic localization. This effect can be further supported by performing a similar EEDL decomposition of the absorption spectrum computed in the gas phase. For the latter, the absorption spectrum is computed from the snapshots of the QM/MM MD simulation where the confining membrane has been removed. The EEDL decomposition of the gas phase spectrum provides contributions of 9%, 55% and 35% for delocalizations over one, two and three ligands, respectively, i.e. the delocalization length of the exciton increases in the absence of the membrane.

In order to specifically determine the effect of the aliphatic C17-chain and the environment on the directionality of the exciton in $[Ru(bpy)_2(bpy-C17)]^{2+}$, we performed a population analysis of the electronic states composing the absorption band by computing the contribution of each of the three ligands to the exciton (Figure 3C). In aqueous solution, the electronic population on the bpy2 ligand (43%) is significantly larger than that on bpy1 (26%) and bpyC17 (31%). In other words, after light absorption, electron transfer from the metal to the bpy2 ligand is the most favorable electronic transition. Why is this ligand preferred? This preference can be attributed to the solvent behavior of the surrounding water molecules. In order to scrutinize this effect, we determined the number of water molecules that solvate each of the three ligands, considering a solvation sphere of 3 Å along the QM/MM MD simulation. Figure 3D shows that, in average, bpy2 is solvated by three water molecules while the bpy1 and bpyC17 ligands are solvated by only two water molecules. This means that the local polarity of the medium is higher around the bpy2 ligand than around bpy1 and bpyC17, and accordingly, electron transfer to byp2 is favored. Such a different local environment is triggered by the presence of the aliphatic chain of bpyC17, which creates a non-polar hole from which the polar water molecules are excluded. Therefore, the ligands closest to the aliphatic chain, bpy1 and bpyC17, are solvated by a smaller number of water molecules than bpy2. When the metal complex is embedded inside the membrane, a large exciton polarization is transferred from bpy2 (34 %) to bpyC17 (39%). This can be explained attending to the binding mode of the complex inside the lipid bilayer. As represented in Figure 2A, when [Ru(bpy)$_2$(bpy-C17)]$^{2+}$ inserts into the membrane, the bpy1 and bpyC17 are in close contact with the surface of the bilayer, which contents the positively charged amino groups of the PC moieties but the bpy2 points towards the solvent. Therefore, after absorption of light, electron transfer from the metal bpyC17 is favored by attractive electrostatic interactions between the excited electron and the positively charged membrane surface.

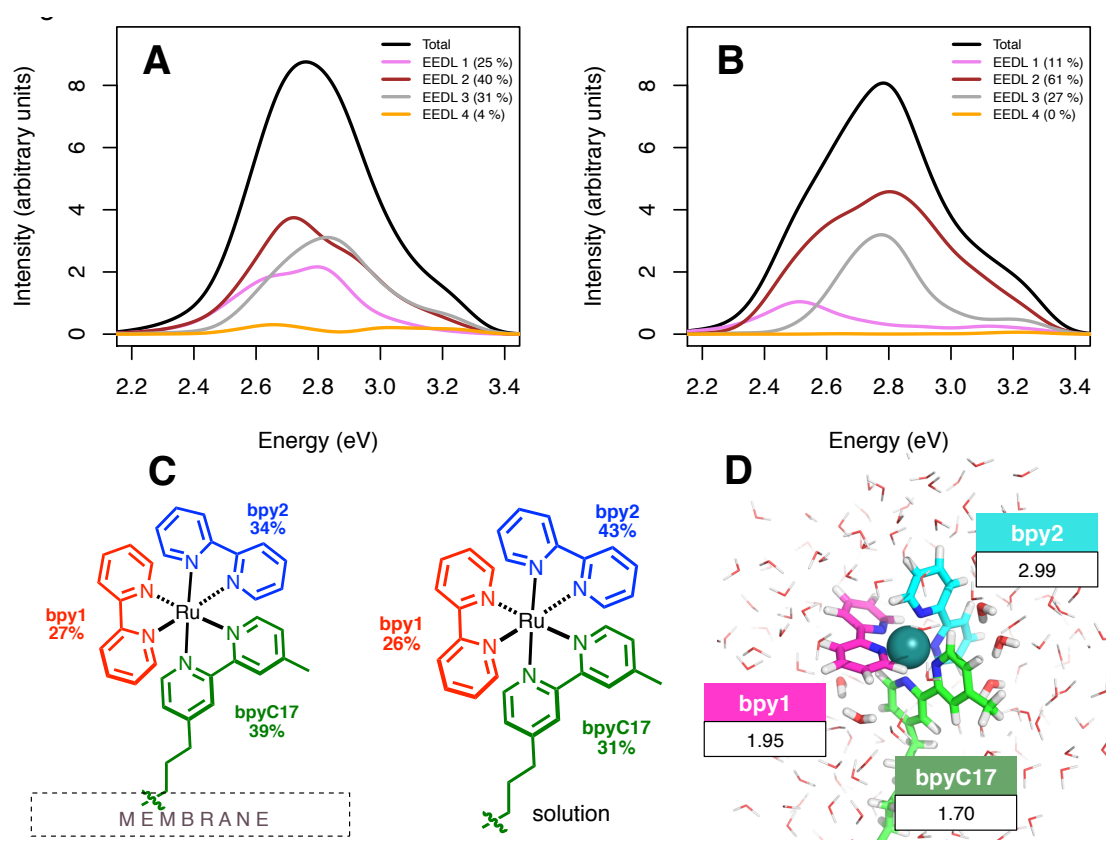

**Figure 3.** Decomposition of the lowest energy band of the UV absorption spectrum for [Ru(bpy)$_2$(bpy-C17)]$^{2+}$ into different excited-electron delocalization length (EEDL) contributions (A) in water and (B) into the membrane. (C) Electronic population analysis of the different ligands in water and into the membrane. (D) Schematic representation of the number of water molecules that solvate the different ligands of the complex along the QM/MM MD dynamics.

In conclusion, we have rationalized a different excited-state electronic behavior of [Ru(bpy)$_2$(bpy-C17)]$^{2+}$ when it is embedded in a lipid membrane or in water. Upon excitation in aqueous solution, the exciton that is created is mostly delocalized over two ligands and the aliphatic chain present on one of the ligands generates a local non-polar environment with low water concentration. As a consequence the excited electron is transferred in the direction opposite to the aliphatic chain, where the local polarity is higher. Instead, when [Ru(bpy)$_2$(bpy-C17)]$^{2+}$ is integrated into the lipid bilayer, the electrostatic interactions with the positively charged amino groups of the membrane induce two different effects: the excited electron is more localized than in water and the electron is transferred preferably to the ligands that are close to the membrane.

Although the photochemical properties of such complexes and its potential applications will also depend on how the initially formed exciton evolves in time, a quantitative analysis of the Franck-Condon excited states in the presence of the environment is a step forward to help the design of Ru(II)-based photosensitizers in biological media. The analysis presented here shows to be sensible to small changes that cannot otherwise be easily detected but that will influence the efficiency of the photosensitizer. This approach can then be used to quantify the effect that functionalization with non-polar substituents or positively charged or electron withdrawing groups will cause in the presence of a particular environment, e.g. lipid membranes. Spatial control of excited electrons, i.e. directional electron transfer between the excited photosensitizer and sacrificial agents, is key in optogenetic experiments as intermolecular electron transfer is more efficient when the separation between the units involved in the process is small. Therefore, localization of the excited electron on a 'selected' moiety of the photosensitizer in close proximity of the sacrificial agent by rational design improves the electron transfer event. Work along these lines is in progress.

ASSOCIATED CONTENT

Computational details are described in detail in the Supporting Information.

AUTHOR INFORMATION

ORCID

Pedro A. Sánchez-Murcia     0000-0001-8415-870X

Juan José Nogueira

Leticia González            0000-0001-5112-794X

ACKNOWLEDGMENT

The University of Vienna is gratefully acknowledged for financial support. P.A.S.M. thanks the Fundación Ramón Areces and the FWF (Project M 2260) for financial support. The Vienna Scientific Cluster (VSC) is also thanked for generous allocation of computer resources.

REFERENCES

1.      Friedman, A. E.; Chambron, J. C.; Sauvage, J. P.; Turro, N. J.; Barton, J. K. A molecular light switch for DNA: Ru(bpy)2(dppz)2+. *J. Am. Chem. Soc.* **1990**, *112* (12), 4960-4962.
2.      Durham, B.; Millett, F. Design of photoactive ruthenium complexes to study electron transfer and proton pumping in cytochrome oxidase. *Biochim. Biophys. Acta (BBA) - Bioenergetics* **2012**, *1817* (4), 567-574.
3.      Bijeire, L.; Elias, B.; Souchard, J.-P.; Gicquel, E.; Moucheron, C.; Kirsch-De Mesmaeker, A.; Vicendo, P. Photoelectron Transfer Processes with Ruthenium(II) Polypyridyl Complexes and Cu/Zn Superoxide Dismutase. *Biochem.* **2006**, *45* (19), 6160-6169.
4.      Rohan, J. G.; Citron, Y. R.; Durrell, A. C.; Cheruzel, L. E.; Gray, H. B.; Grubbs, R. H.; Humayun, M.; Engisch, K. L.; Pikov, V.; Chow, R. H. Light-Triggered Modulation of Cellular Electrical Activity by Ruthenium Diimine Nanoswitches. *Acs Chem Neurosci* **2013**, *4* (4), 585-593.
5.      Preiß, J.; Jäger, M.; Rau, S.; Dietzek, B.; Popp, J.; Martínez, T.; Presselt, M. How Does Peripheral Functionalization of Ruthenium(II)–Terpyridine Complexes Affect Spatial Charge Redistribution after Photoexcitation at the Franck–Condon Point? *ChemPhysChem* **2015**, *16* (7), 1395-1404.
6.      Jäger, M.; Freitag, L.; González, L. Using computational chemistry to design Ru photosensitizers with directional charge transfer. *Coord. Chem. Rev.* **2015**, *304*, 146-165.

7. Yeh, A. T.; Shank, C. V.; McCusker, J. K. Ultrafast Electron Localization Dynamics Following Photo-Induced Charge Transfer. *Science* **2000,** *289* (5481), 935-938.
8. Chergui, M. Ultrafast Photophysics of Transition Metal Complexes. *Acc. Chem. Res.* **2015,** *48* (3), 801-808.
9. Marquetand, P.; Nogueira, J. J.; Mai, S.; Plasser, F.; González, L. Challenges in simulating light-induced processes in DNA. *Molecules* **2017,** *22* (1), 49.
10. Nogueira, J. J.; Meixner, M.; Bittermann, M.; González, L. Impact of Lipid Environment on Photodamage Activation of Methylene Blue. *ChemPhotoChem* **2017,** *1*, 178–182.
11. Nogueira, J. J.; Plasser, F.; González, L. Electronic delocalization, charge transfer and hypochromism in the UV absorption spectrum of polyadenine unravelled by multiscale computations and quantitative wavefunction analysis. *Chem. Sci.* **2017,** *8* (8), 5682-5691.
12. Dongare, P.; Myron, B. D. B.; Wang, L.; Thompson, D. W.; Meyer, T. J. [Ru(bpy)3]2+∗ revisited. Is it localized or delocalized? How does it decay? *Coord. Chem. Rev.* **2017,** *345*, 86-107.
13. Damrauer, N. H.; Cerullo, G.; Yeh, A.; Boussie, T. R.; Shank, C. V.; McCusker, J. K. Femtosecond Dynamics of Excited-State Evolution in [Ru(bpy)3]2+. *Science* **1997,** *275* (5296), 54-57.
14. McCusker, J. K. Femtosecond Absorption Spectroscopy of Transition Metal Charge-Transfer Complexes. *Acc. Chem. Res.* **2003,** *36* (12), 876-887.
15. Cannizzo, A.; van Mourik, F.; Gawelda, W.; Zgrablic, G.; Bressler, C.; Chergui, M. Broadband Femtosecond Fluorescence Spectroscopy of [Ru(bpy)3]2+. *Angew. Chem. Int. Ed.* **2006,** *45* (19), 3174-3176.
16. Strouse, G. F.; Schoonover, J. R.; Duesing, R.; Boyde, S.; Jones, W. E., Jr.; Meyer, T. J. Influence Of Electronic Delocalization In Metal-to-Ligand Charge Transfer Excited States. *Inorg. Chem.* **1995,** *34* (2), 473-487.
17. Kober, E. M.; Sullivan, B. P.; Meyer, T. J. Solvent dependence of metal-to-ligand charge-transfer transitions. Evidence for initial electron localization in MLCT excited states of 2,2'-bipyridine complexes of ruthenium(II) and osmium(II). *Inorg. Chem.* **1984,** *23* (14), 2098-2104.
18. Atkins, A. J.; González, L. Trajectory Surface-Hopping Dynamics Including Intersystem Crossing in [Ru(bpy)3]2+. *J. Phys. Chem. Lett.* **2017,** *8* (16), 3840-3845.
19. Brandt, P.; Norrby, T.; Åkermark, B.; Norrby, P.-O. Molecular Mechanics (MM3*) Parameters for Ruthenium(II)−Polypyridyl Complexes. *Inorg. Chem.* **1998,** *37* (16), 4120-4127.
20. Moret, M.-E.; Tavernelli, I.; Rothlisberger, U. Combined QM/MM and Classical Molecular Dynamics Study of [Ru(bpy)3]2+ in Water. *J. Phys. Chem. B* **2009,** *113* (22), 7737-7744.
21. Wilson, T.; Costa, P. J.; Félix, V.; Williamson, M. P.; Thomas, J. A. Structural Studies on Dinuclear Ruthenium(II) Complexes That Bind Diastereoselectively to an Antiparallel Folded Human Telomere Sequence. *J. Med. Chem.* **2013,** *56* (21), 8674-8683.
22. Dickson, C. J.; Madej, B. D.; Skjevik, Å. A.; Betz, R. M.; Teigen, K.; Gould, I. R.; Walker, R. C. Lipid14: The Amber Lipid Force Field. *J. Chem. Theory Comput.* **2014,** *10* (2), 865-879.
23. Jorgensen, W. L.; Chandrasekhar, J.; Madura, J. D.; Impey, R. W.; Klein, M. L. Comparison of simple potential functions for simulating liquid water. *J. Chem. Phys.* **1983,** *79* (2), 926-935.
24. van Meer, G.; Voelker, D. R.; Feigenson, G. W. Membrane lipids: where they are and how they behave. *Nat. Rev, Mol. Cell. Biol.* **2008,** *9* (2), 112-124.
25. Kristyna, P.; Rainer, A. B. Biomembranes in atomistic and coarse-grained simulations. *J. Phys. Cond. Mat.* **2015,** *27* (32), 323103.
26. Becke, A. D. A new mixing of Hartree‑Fock and local density‑functional theories. *J. Chem. Phys.* **1993,** *98* (2), 1372-1377.
27. Juris, A.; Balzani, V.; Barigelletti, F.; Campagna, S.; Belser, P.; von Zelewsky, A. Ru(II) polypyridine complexes: photophysics, photochemistry, eletrochemistry, and chemiluminescence. *Coord. Chem. Rev.* **1988,** *84*, 85-277.
28. Plasser, F.; Lischka, H. Analysis of excitonic and charge transfer interactions from quantum chemical calculations. *J. Chem. Theory Comput.* **2012,** *8* (8), 2777-2789.
29. Plasser, F.; Wormit, M.; Dreuw, A. New tools for the systematic analysis and visualization of electronic excitations. I. Formalism. *J. Chem. Phys.* **2014,** *141* (2), 024106.
30. Mai, S.; Gattuso, H.; Fumanal, M.; Munoz-Losa, A.; Monari, A.; Daniel, C.; Gonzalez, L. Excited-states of a rhenium carbonyl diimine complex: solvation models, spin-orbit coupling, and vibrational sampling effects. *Phys. Chem. Chem. Phys.* **2017,** *19* (40), 27240-27250.

31. Moret, M.-E.; Tavernelli, I.; Chergui, M.; Rothlisberger, U. Electron Localization Dynamics in the Triplet Excited State of [Ru(bpy)3]2+ in Aqueous Solution. *Chem. Eur. J.* **2010,** *16* (20), 5889-5894.